\documentclass[aps,prl,showpacs,amsmath,amssymb,superscriptaddress,longbibliography,reprint,10pt]{revtex4-2}
\usepackage{graphicx}
\usepackage{xcolor}
\usepackage{newtxtext,newtxmath}
\usepackage{mathrsfs} 
\usepackage{braket}

\usepackage[colorlinks=true,breaklinks=true,allcolors=blue]{hyperref}

\definecolor{michael}{rgb}{.2,.5,.6}

\definecolor{hannes}{rgb}{1,0,0}

\definecolor{hannesc}{rgb}{1,0,1}

\definecolor{emma}{rgb}{0,0.5,0}

\begin{document}

\title{Universal Cooling Dynamics Toward a Quantum Critical Point}

\author{Emma C. King}
\affiliation{Institute of Theoretical Physics, Stellenbosch University, Stellenbosch 7600, South Africa}

\author{Johannes N. Kriel}
\affiliation{Institute of Theoretical Physics, Stellenbosch University, Stellenbosch 7600, South Africa}

\author{Michael Kastner}
\affiliation{Institute of Theoretical Physics, Stellenbosch University, Stellenbosch 7600, South Africa}
\affiliation{Hanse-Wissenschaftskolleg, Lehmkuhlenbusch 4, 27753 Delmenhorst, Germany}

\date{\today}

\begin{abstract}
We investigate the loss of adiabaticity when cooling a many-body quantum system from an initial thermal state toward a quantum critical point. The excitation density, which quantifies the degree of adiabaticity of the dynamics, is found to obey scaling laws in the cooling velocity as well as in the initial and final temperatures of the cooling protocol. The scaling laws are universal, governed by the critical exponents of the quantum phase transition. The validity of these statements is shown analytically for a Kitaev quantum wire coupled to Markovian baths and argued to be valid under rather general conditions. Our results establish that quantum critical properties can be probed dynamically at finite temperature, without even varying the control parameter of the quantum phase transition.
\end{abstract}


\maketitle 

Critical phenomena, scaling laws, and universality are key concepts in equilibrium physics of many-body systems. Extending these concepts to out-of-equilibrium situations is a key challenge and a vibrant research field, aiming at consolidating our understanding of nonequilibrium many-body systems. Efforts in this direction include nonequilibrium phase transitions in classical stochastic dynamics \cite{MarroDickman}, transitions in trajectory space \cite{Hedges_etal09}, quantum phase transitions of nonequilibrium steady states \cite{ProsenPizorn09}, as well as dynamical quantum phase transitions of various kinds \cite{EcksteinKollarWerner09,*Diehl_etal09,*HeylPolkovnikovKehrein13,*Halimeh_etal17}. In addition to such genuine nonequilibrium transitions, a related research direction investigates the imprint of equilibrium phase transitions onto a system's nonequilibrium dynamics, which may lead to the emergence of universal scaling laws out of equilibrium. Such imprints can also be used to probe equilibrium physics by means of nonequilibrium protocols, which may be useful in experimental situations where equilibrium is difficult to reach \cite{Karl_etal17,*DagSun21}.

\begin{figure}[b]\centering%
\includegraphics[width=0.78\linewidth]{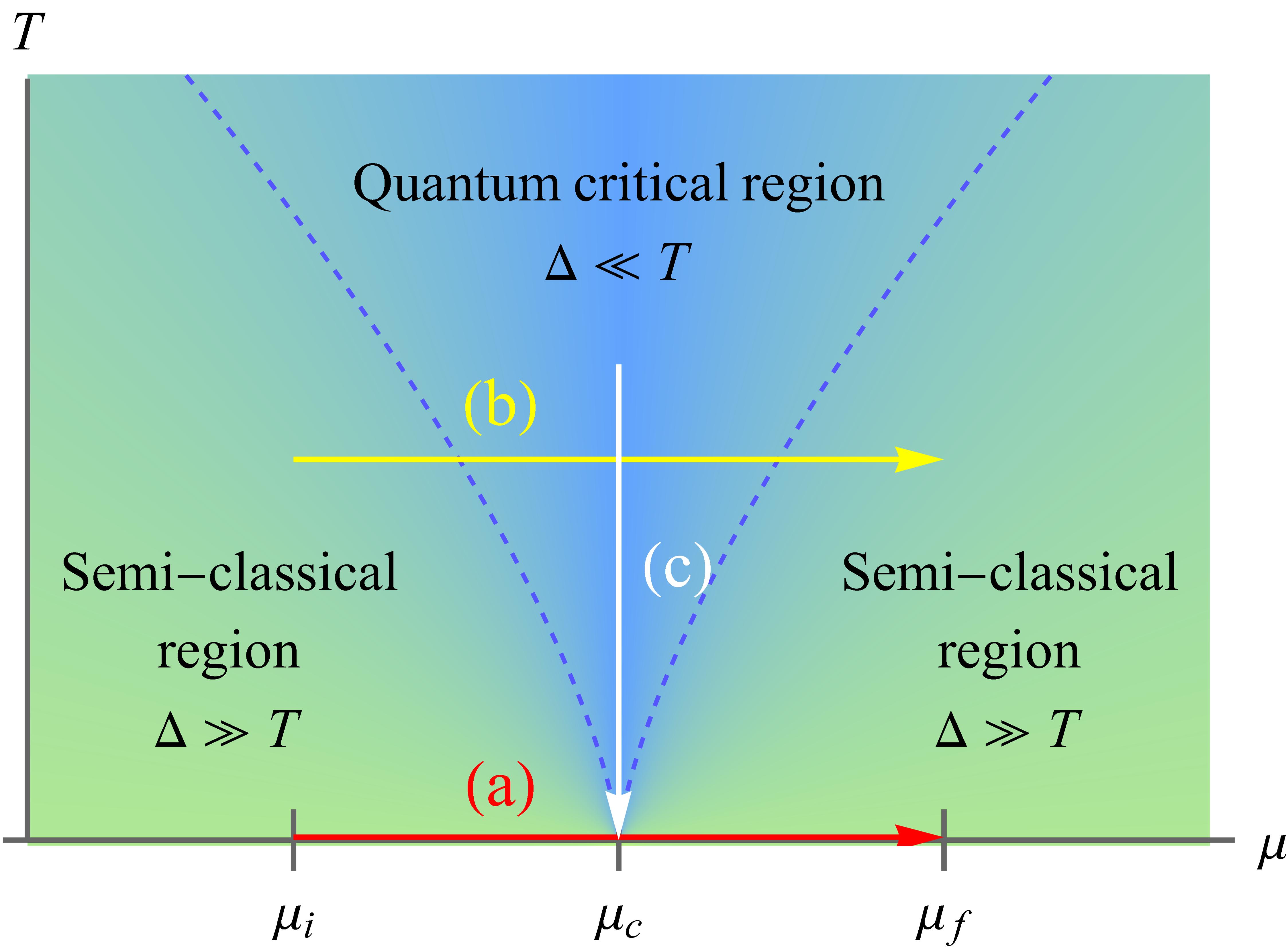}%
\caption{\label{f:rampingdiagram}%
Ramping protocols in the vicinity of a quantum critical point, with $\mu$ a control parameter in the Hamiltonian. (a) Standard quantum Kibble-Zurek protocol at temperature $T=0$, involving a linear parameter ramp from the initial value $\mu_i$ to the final value $\mu_f$, crossing the critical point $\mu=\mu_c$. (b) As in (a), but at constant positive temperature. (c) Cooling from an initial temperature $T_i>0$ to the final temperature $T_f\simeq0$ with $\mu$ fixed at $\mu_c$.%
}%
\end{figure}

A prominent manifestation of equilibrium criticality under nonequilibrium conditions goes under the name of {\em Kibble-Zurek mechanism}. Initially proposed by Kibble to explain domain formation in the early Universe \cite{Kibble76,*Kibble80}, and subsequently extended by Zurek to continuous phase transitions in condensed matter systems \cite{Zurek85,*Zurek96}, the Kibble-Zurek mechanism is a consequence of {\em critical slowing down}, i.e., the power law divergence of the relaxation time of a many-body system in the vicinity of a continuous phase transition. For a system in equilibrium at some initial temperature $T$ sufficiently far from the phase transition temperature $T_c$, a gradual change of $T$ results in adiabatic dynamics and leaves the system equilibrated. Only when $T$ gets sufficiently close to $T_c$, critical slowing down prevents further adiabatic evolution and causes an approximate freeze-out in a nonequilibrium state that encodes signatures of the equilibrium phase transition. An extension of this thermal Kibble-Zurek mechanism to quantum phase transitions at zero temperature was proposed in later works \cite{ZurekDornerZoller05,*Polkovnikov05}. Here, a quantum many-body system is prepared in its ground state, whereupon a parameter in the Hamiltonian is slowly ramped toward (and possibly across) its critical value, producing universal signatures in the excitations generated once critical slowing down causes adiabaticity to breakdown; see (a) in Fig.~\ref{f:rampingdiagram}. Experimental verifications, or at least consistency checks, of Kibble-Zurek physics have been reported in the past decade \cite{delCampoZurek14,*Anquez_etal16,*BeugnonNavon17}. Generalizations of Kibble-Zurek physics to open systems are known, but tend to suffer from the presence of multiple timescales, which in turn lead to complicated crossovers that obfuscate clean scaling behavior \cite{Patane_etal08,*Patane_etal09,Nalbach_etal15,*DuttaRahmaniDelCampo16}.

In this Letter, we report clean scaling and universality in open nonequilibrium quantum systems that are cooled toward a quantum critical point; see (c) in Fig.~\ref{f:rampingdiagram}. Conceptually, this setting combines the temperature ramps of Kibble's original proposal with the exploration of quantum critical phenomena. Even though the transition between equilibrium quantum phases is not temperature driven, but occurs under variation of a Hamiltonian parameter, we show that nonequilibrium properties of a system that is cooled toward its quantum critical point carry an imprint of the equilibrium quantum critical exponents. Accordingly, equilibrium critical exponents of a zero-temperature quantum phase transition can be probed in a nonequilibrium setting at finite temperature without even varying the external parameter that drives the equilibrium transition. In technical terms, we derive universal scaling functions for the density of excess excitations that survive the cooling process, a quantity that measures the degree to which adiabaticity is violated. The excess excitation density scales as a power law in the cooling velocity, the initial temperature, and also the final temperature of the cooling protocol, with scaling exponents that depend on the equilibrium quantum critical exponents and the spectral density exponent of the bath.

The widespread use of cooling processes in physics and the recent experimental progress in cooling quantum many-body systems toward their quantum critical regime, make a case for the relevance and timeliness of identifying universal features of cooling processes. A variety of experiments should be suitable for verifying our findings in principle, including ultracold atoms as well as solid-state experiments at low temperatures. Understanding the defect creation under variation of the temperature close to a quantum critical point may also prove beneficial for the design of adiabatic quantum computation protocols with controlled dissipation.

{\em Kitaev quantum wire in a thermal bath.---}To observe universality under temperature ramps in a many-body open quantum system, we require that
\makeatletter
\let\orig@listi\@listi
\def\@listi{\orig@listi\topsep=0\baselineskip}
\makeatother
\begin{enumerate}
\setlength{\itemsep}{0pt}
\setlength{\parskip}{0pt}
\item the corresponding closed system undergoes an equilibrium quantum phase transition,\label{i:1}
\item it thermalizes under the open-system time evolution, and\label{i:2}
\item the open-system dynamics is exactly solvable, allowing us to treat large system sizes.\label{i:3}
\end{enumerate}
We consider a Kitaev chain \cite{Kitaev01} of $L$ sites with Hamiltonian
\begin{equation}\label{e:Kitaev}
H \!=\! \sum_{i=1}^L \left[J\left( c_i^\dagger c_{i+1}^{\vphantom{\dagger}} + c_{i+1}^\dagger c_i^{\vphantom{\dagger}}\right) + \frac{\Delta}{2}\left( c_i^{\vphantom{\dagger}} c_{i+1}^{\vphantom{\dagger}} + c_{i+1}^\dagger c_i^\dagger\right) + 2 \mu c_i^\dagger c_i^{\vphantom{\dagger}} \right]\!,
\end{equation}
where $c_i$ denotes a spinless fermionic operator acting on site $i$. The parameters $J$ and $\Delta$ denote the hopping and pairing strengths, and $\mu$ is the chemical potential. We impose periodic boundary conditions. The Hamiltonian \eqref{e:Kitaev}, being quadratic in the fermionic operators, can be diagonalized by Fourier and Bogoliubov transformations, yielding $H=\sum_k \lambda_k \eta^\dag_k \eta_k$, with $\eta_k$ the Bogoliubov fermionic operators and $\lambda_k$ the mode energies; see \cite{KingKastnerKriel} for details. The Kitaev chain undergoes a quantum phase transition between topologically distinct phases at $\mu=\pm J$ \cite{Kitaev01} and hence satisfies condition \ref{i:1}.

To derive an open-system master equation that thermalizes at late times, we consider $L$ identical and independent bosonic baths, each of which is weakly coupled to one of the sites of the Kitaev chain \footnote{A translationally invariant bath configuration is convenient, but not a requirement for what follows.}. Using results by D'Abbruzzo and Rossini \cite{DAbbruzzoRossini21}, a Markovian master equation in Lindblad form can be derived in a self-consistent way,
\begin{equation}\label{e:Lindblad}
\dot{\rho} = -i\left[H,\rho\right] + \gamma \sum_k \sum_{\sigma=\pm}\left(2 L_{k\sigma}^{\phantom{\dagger}}\rho L_{k\sigma}^\dagger - \left\{L_{k\sigma}^\dagger L_{k\sigma}^{\phantom{\dagger}},\rho\right\} \right),
\end{equation}
where $\rho$ denotes the density operator of the Kitaev chain, $\gamma$ is the system--bath coupling, and angular and curly brackets denote commutators and anticommutators, respectively. The jump operators $L_{k\pm}$ are of the form $L_{k+}=\sqrt{\Gamma_{k+}}\eta_k^\dag$ and $L_{k-}=\sqrt{\Gamma_{k-}}\eta_k$, where the couplings $\Gamma_{k\pm}$ contain the baths' temperatures and spectral densities. A derivation of Eq.\ \eqref{e:Lindblad} and expressions for the jump operators are given in the companion paper \cite{KingKastnerKriel}. Unlike in Kitaev chains with {\em ad hoc} introduced dissipation \cite{Keck_etal17,*vanCaspel_etal19,*RossiniVicari20}, we ended up with jump operators that are nonlocal in the lattice fermions $c_i$, which is essential for thermalization to occur and condition \ref{i:2} to be satisfied.

The jump operators $L_{k\pm}$ are linear in the Bogoliubov fermions $\eta_k$; see \cite{KingKastnerKriel} for details. As a result, the master equation \eqref{e:Lindblad} is quadratic in the fermionic operators and can be diagonalized in Liouville space by the method of third quantization \cite{Prosen08}, which accounts for property \ref{i:3} of the above list. For the analysis of universal features and the loss of adiabaticity of the dynamics of this model, the main quantities of interest are the time-dependent mode occupation numbers $\mathcal{P}_k=\braket{\eta_k^\dagger \eta_k^{\phantom{\dagger}}}$. These in turn can be expressed in terms of two-point correlation functions in Liouville space, which we calculate by a formalism due to Kos and Prosen \cite{KosProsen17}. Calculating $\mathcal{P}_k$ then amounts to numerically solving $4\times4$-matrix differential equations with time-dependent coefficients, as reported in detail in the companion paper \cite{KingKastnerKriel}.

\begin{figure*}\centering
\includegraphics[width=0.32\linewidth]{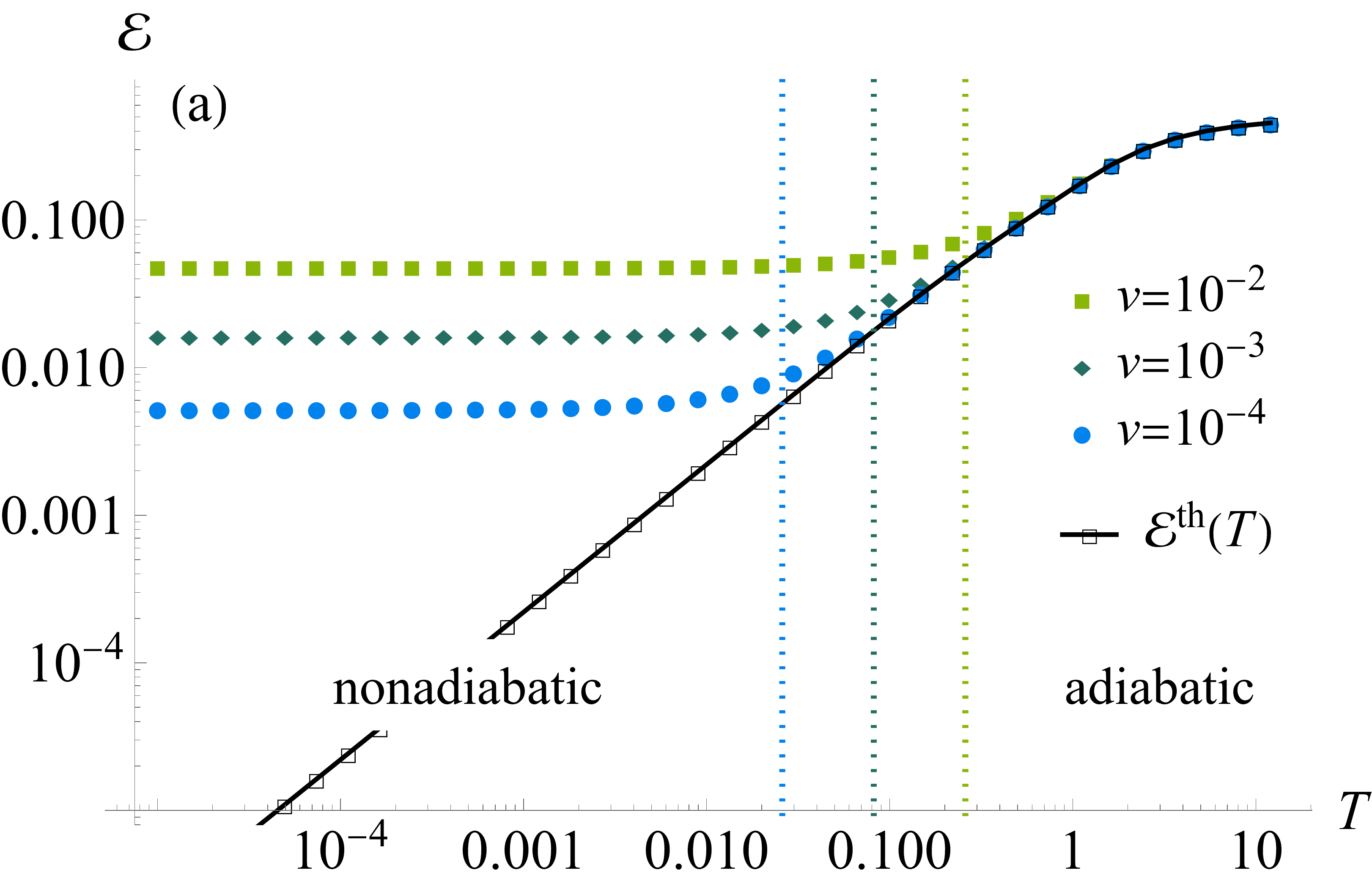}
\includegraphics[width=0.32\linewidth]{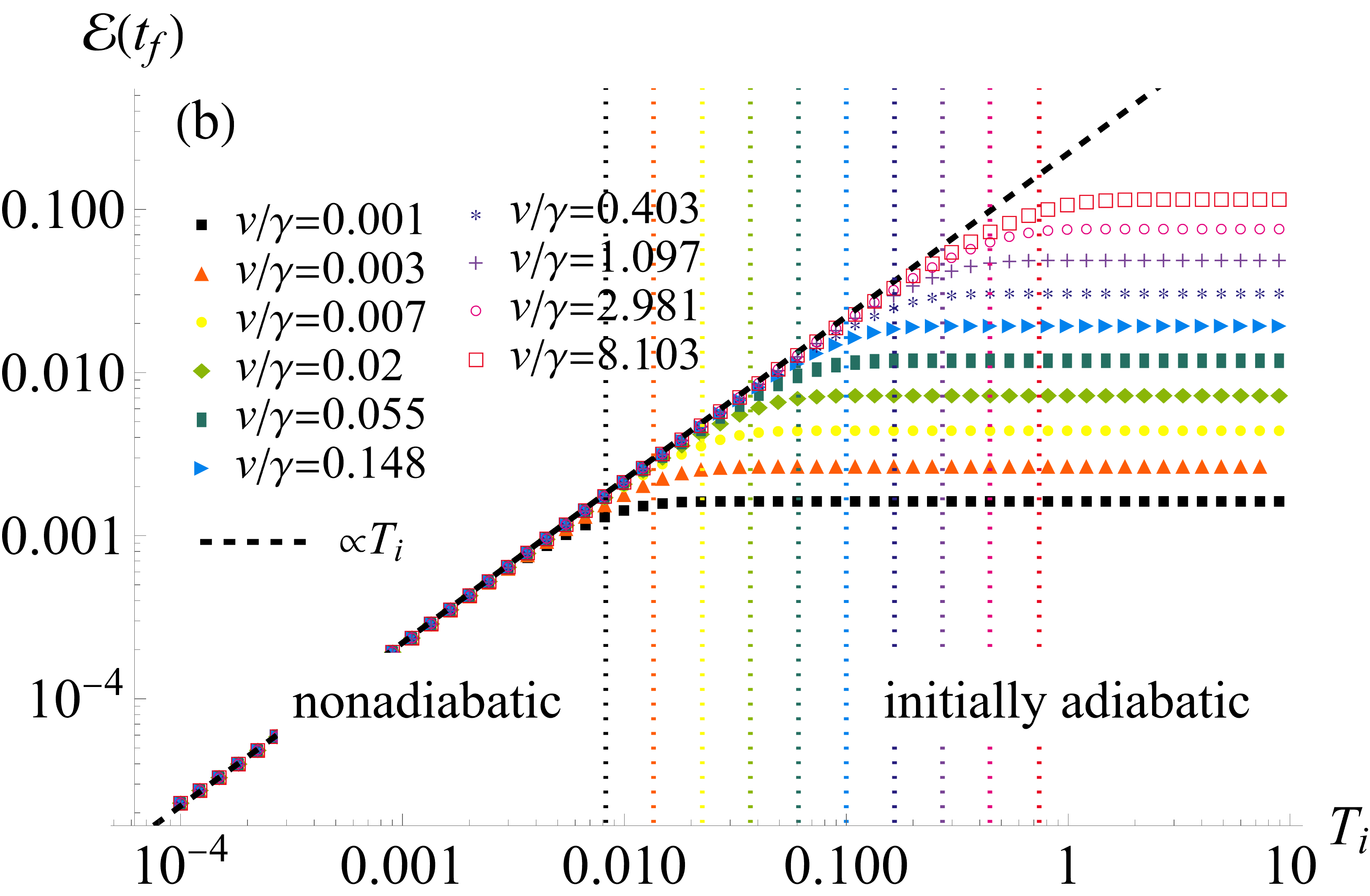}
\includegraphics[width=0.32\linewidth]{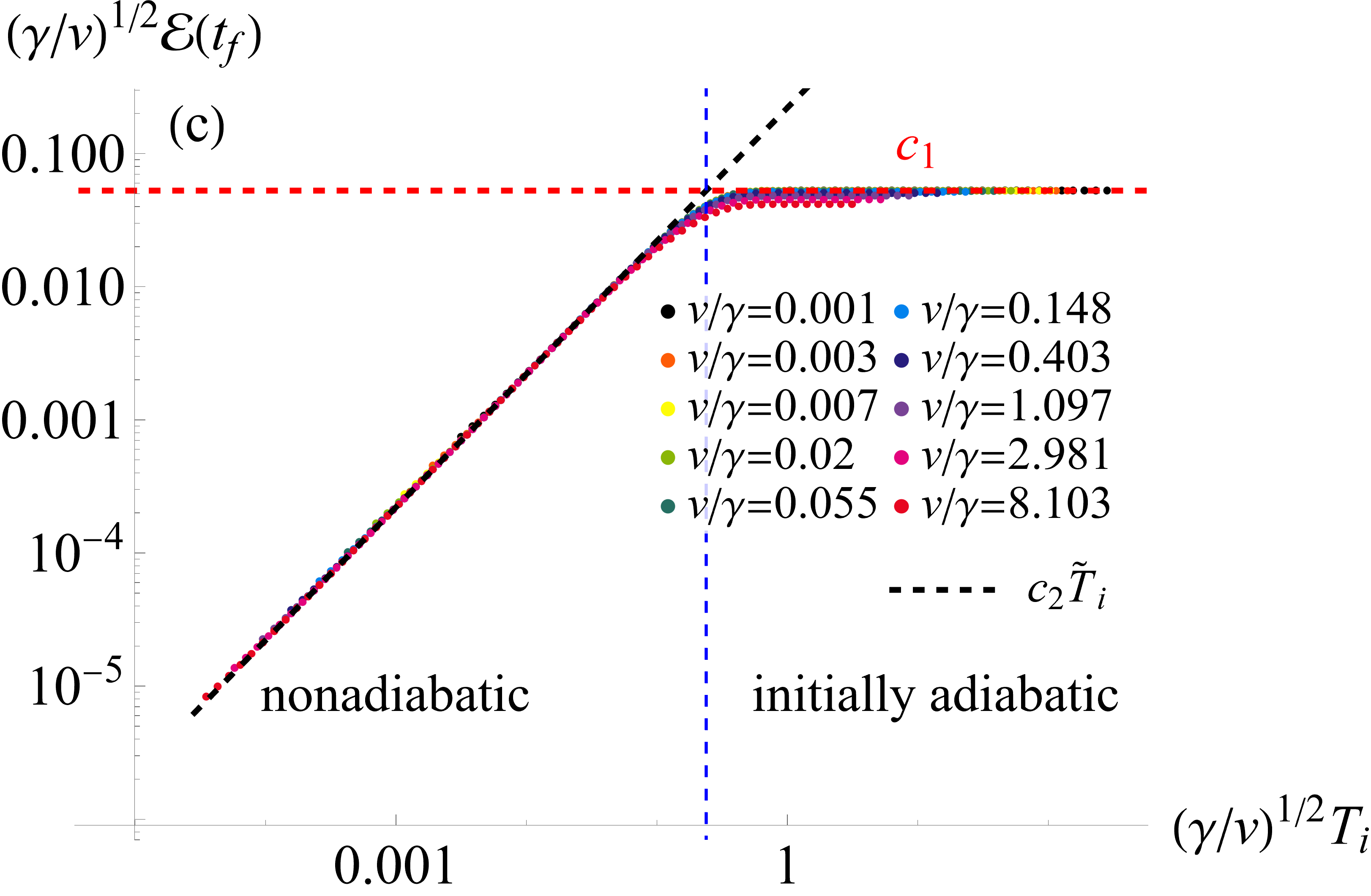}
\caption{\label{f:E(t)}%
(a) Excitation densities $\mathcal{E}$ {\em vs.}\ $T(t)$ for various ramp velocities $v$, calculated by numerically solving the rate equations \eqref{e:rateequation} and inserting $\mathcal{P}_k$ into 
\eqref{e:excitationdensity}. The dynamics starts from a thermal state at $T_i=15$ (top right corner). Dotted lines in the corresponding colors mark the crossover temperatures between adiabatic and nonadiabatic cooling. Parameter values are $L=10^6$, $J=\Delta=1$, $\delta=1$, $\mu=-1$, $\gamma=0.01$. Excitation densities $\mathcal{E}^\text{th}$ of equilibrium distributions with temperature $T(t)$ are shown in black. (b)~Excitation densities $\mathcal{E}(t_f)$ at the end of the ramps as functions of the initial temperature $T_i$, shown for various values of $v/\gamma$. Parameters are $L=4\times 10^4$, $J=\Delta=1$, $\delta=1$, $\mu=-1$. Dotted vertical lines mark the crossover between adiabatic and nonadiabatic regimes. The dashed black line $\propto T_i$ marks the limit $v\to\infty$, in which the dynamics is nonadiabatic from the outset for any $T_i$. (c) As in (b), but showing the rescaled excitation density $\mathcal{E}(t_f)\sqrt{\gamma/v}$ {\em vs.}\ the rescaled initial temperature $\widetilde{T}_i=T_i\sqrt{\gamma/v}$, resulting in approximate data collapse onto a single curve. 
}%
\end{figure*}

{\em Temperature ramps in the Kitaev chain.---}When deriving the master equation \eqref{e:Lindblad}, the bath temperature $T$ of the microscopic model gets imprinted onto the jump operators $L_{k\sigma}$, which become $T$ dependent. To realize the cooling protocol (c) illustrated in Fig.~\ref{f:rampingdiagram}, we consider $T$ as a time-dependent quantity that decreases linearly with velocity $v$ from some initial value $T_i>0$ to zero, $T(t)=T_i - vt$ with $t\in[0,t_f=T_i/v]$. This renders the right-hand side of Eq.\ \eqref{e:Lindblad} explicitly time dependent. The ramp dynamics starts from a thermalized initial state $\rho\propto\exp(-H/T_i)$ at $t=0$. If the ramp was carried out infinitely slowly, i.e., with $v\simeq0$, then the system would evolve through a continuum of thermal states with temperatures $T(t)$ for the entire duration of the ramping protocol. For finite velocities $v$, however, adiabaticity is violated and excess excitations survive on top of the thermal excitations of an equilibrium state with temperature $T(t)$. 

When specializing to temperature ramps, the earlier mentioned matrix differential equations for mode occupations $\mathcal{P}_k$ can be simplified, resulting in uncoupled rate equations 
\begin{equation}\label{e:rateequation}
\frac{d}{dt}\mathcal{P}_k = -\frac{1}{\tau(\lambda_k,T(t))}\left\{\mathcal{P}_k-\mathcal{P}^\text{th}[\lambda_k/T(t)]\right\},
\end{equation}
see \cite{KingKastnerKriel} for a derivation. Here, $\lambda_k$ denotes the energy of the $k$-mode and $\mathcal{P}^\text{th}(x)=[\exp(x)+1]^{-1}$ is the Fermi-Dirac distribution. The mode relaxation rate
\begin{equation}\label{e:rate}
\tau^{-1}(\lambda_k,T)=2\gamma\mathcal{J}(\lambda_k)\coth[\lambda_k/(2T)]
\end{equation}
is proportional to the system--bath coupling strength $\gamma$. The bath spectral density $\mathcal{J}$ of the bosonic baths from which the Lindblad equation \eqref{e:Lindblad} was derived is arbitrary, but we usually consider power law spectral densities $\mathcal{J}(\lambda)=\pi\delta\lambda^s\exp(-\lambda/\lambda_c)$ with parameter $\delta$. In the regime where the Markov approximation is justified, the cutoff frequency $\lambda_c$ can be chosen well above the mode frequencies without loss of generality, allowing for the approximation $\mathcal{J}(\lambda_k)\approx\pi\delta\lambda_k^s$. Based on the rate equations \eqref{e:rateequation}, the time evolution of mode occupation numbers can be calculated for temperature ramps in fairly large systems of $10^6$ lattice sites and more.

To measure the degree to which adiabaticity is violated in the course of such ramps, we use the excitation density
\begin{equation}\label{e:excitationdensity}
\mathcal{E}(t)=\frac{1}{L}\sum_k \mathcal{P}_k(t) = \frac{1}{L}\sum_k \braket{\eta_k^\dagger \eta_k^{\phantom{\dagger}}}(t).
\end{equation}
Figure \ref{f:E(t)}(a) shows plots of the time evolution of the excitation density \eqref{e:excitationdensity} of the Kitaev chain \eqref{e:Kitaev} at the critical parameter value $\mu=-J=-1$ for temperature ramps of different velocities. At the initial time $t=0$ the system is thermalized and the excitation density $\mathcal{E}(0)$ agrees with the excitation density $\mathcal{E}^\text{th}$ of the thermal equilibrium distribution with temperature $T(0)=T_i$ [top right corner of Fig.~\ref{f:E(t)}(a)]. During the initial phase of the ramp, the system thermalizes fast, hence cools down concurrently with the bath such that the excitation density agrees with the thermal one. The closer the quantum critical point at $T=0$ is approached, the longer thermalization takes, until adiabaticity breaks down and excess excitations survive in addition to the thermal ones [left part of Fig.~\ref{f:E(t)}(a)]. The point at which adiabaticity is lost depends on the ramp velocity $v$ (as indicated by vertical lines) and the initial temperature $T_i$ [Fig.~S1(a) of the Supplemental Material \footnote{See Supplemental Material [url], which includes Refs.\ \cite{Dutta_etal,NIST,Vodola_etal16,FisherMaNickel72,
DuttaBhattacharjee01,Defenu_etal20,Viyuela_etal16,BhattacharyaDutta18,
Defenu_etal19}.}].

In this crossover region, $\mathcal{E}$ ``freezes'' at an approximately constant value, which gives rise to the plateaus in Fig.~\ref{f:E(t)}(a). The plateau heights therefore encode information about the value of $T$ at which adiabaticity was lost. For a condensed representation of the ramp data, we plot the plateau values $\mathcal{E}(t_f)$, where $T(t_f)=0$, as a function of the initial temperature $T_i$ for various ramp velocities $v$ [Fig.~\ref{f:E(t)}(b)]. Each point in the plot represents an entire ramp protocol. Points on the dashed black diagonal correspond to ramps with sufficiently small $T_i$ and large $v$ such that adiabaticity is lost immediately when the protocol is started. Points below the dashed black diagonal correspond to ramps where initially the dynamics is adiabatic, followed by a nonadiabatic evolution at a later stage. The flat plateaus on the right-hand side of the plot indicate that adiabaticity is lost at a fairly sharp freeze-out point, roughly at the same temperature $T(t)$, independent of the initial temperature $T_i$. This is good news for our aim of deriving a scaling theory \`a la Kibble-Zurek, for which the sharp separation of adiabatic and frozen regimes is a presupposition. In fact, the separate curves in Fig.~\ref{f:E(t)}(b) can be made to collapse onto each other by rescaling the plot axes with suitable powers of $v$; see Fig.~\ref{f:E(t)}(c). Similar behavior, albeit with different scaling powers, is found for $\mathcal{E}$ as a function of $v$ \cite{Note2}. These findings suggest that the excitation density $\mathcal{E}$ obeys scaling laws with respect to both $T_i$ and $v$.

{\em Scaling theory for temperature ramps.---}To understand the observed data collapse, we use the rate equation \eqref{e:rateequation} as a starting point for deriving a scaling theory for the excitation density under temperature ramps. Despite explicit time dependencies in the relaxation rate \eqref{e:rate} and the thermal equilibrium distribution, Eq.\ \eqref{e:rateequation} can be solved analytically (see \cite{PolyaninZaitsev}, Sec.~1.1.4). Assuming large system sizes and restricting to moderate excitation densities, the result of the calculation is
\begin{equation}\label{e:ExcitationsIntegrated}
\mathcal{E}(T,T_i,\gamma/v) =\frac{1}{\pi z c^{1/z}} \int_0^\infty\! d\lambda\, \lambda^{1/z-1} \mathcal{P}\left(\frac{T}{\lambda},\frac{T_i}{\lambda},\frac{\gamma\lambda^{s +1}}{v}\right)
\end{equation}
with
\begin{subequations}
\begin{align}
\mathcal{P}(x,y,z) &= \mathcal{P}^\text{th}(1/y)e^{z f(x,y)} - 2\pi\delta z \label{e:Pfunction2}\\
&\times \int_y^{x} dx' e^{-z f(x',x)}
\coth[1/(2x')]\mathcal{P}^\text{th}(1/x'),\nonumber\\
f(x,y) &= 2\pi\delta \int_y^{x} dx''
\coth[1/(2x'')],\label{e:F2}
\end{align}
\end{subequations}
see Supplemental Material \cite{Note2} for a derivation. The equilibrium dynamical critical exponent $z$ and constant $c$ in Eq.\ \eqref{e:ExcitationsIntegrated} are determined by the leading order expansion $\lambda_k=c|k|^z$ of the dispersion relation of the Hamiltonian at the critical point, yielding $c=|\Delta|$ and $z=1$ for the Kitaev chain \eqref{e:Kitaev}. The scaling plot in Fig.~\ref{f:E(t)}(c) suggests that Eqs.\ \eqref{e:ExcitationsIntegrated}--\eqref{e:F2} possess an inherent structure. Indeed, it is straightforward to verify that $\mathcal{E}$ is a generalized homogeneous function,
\begin{equation}\label{e:homogeneous}
\mathcal{E}\bigl(\ell^z T,\ell^z T_i,\ell^{-z(s+1)}\gamma/v\bigr) = \ell\, \mathcal{E}(T,T_i,\gamma/v)
\end{equation}
for arbitrary $\ell$. Specializing to ramps ending at $T=0$ and choosing $\ell=(\gamma/v)^{1/[z(s+1)]}$ yields
\begin{equation}\label{e:tildeE1}
(\gamma/v)^{1/[z(s+1)]}\mathcal{E}(0,T_i,\gamma/v) = \mathcal{E}(0,\widetilde{T}_i,1),
\end{equation}
which demonstrates that a properly rescaled excitation density $\mathcal{E}$ is a function of a single variable $\widetilde{T}_i\equiv(\gamma/v)^{1/(s+1)}T_i$ only. Put differently, numerical evaluation of the univariate function on the right-hand side of Eq.\ \eqref{e:tildeE1} gives access to the bivariate function $\mathcal{E}(0,T_i,\gamma/v)$. Inserting the dynamical critical exponent $z=1$ of the Kitaev chain and $s=1$ for an Ohmic spectral density, this result confirms and explains the data collapse observed in Fig.~\ref{f:E(t)}(c). Small imperfections of the numerical data collapse may be attributed to the idealizing assumptions made in the derivation of the analytic results \eqref{e:ExcitationsIntegrated}--\eqref{e:F2}. An asymptotic analysis of Eqs.\ \eqref{e:ExcitationsIntegrated}--\eqref{e:F2}, detailed in the Supplemental Material \cite{Note2}, reproduces the constant behavior $(\gamma/v)^{1/2}\mathcal{E}\sim c_1$ in the limit of large $\widetilde{T}_i$ observed in Fig.~\ref{f:E(t)}c, as well as the linear increase $(\gamma/v)^{1/2}\mathcal{E}\sim c_2\widetilde{T}_i^{1/z}$ for small $\widetilde{T}_i$, with constants $\lvert\Delta\rvert c_1=\int_0^\infty d\lambda\mathcal{P}(0,\infty,\lambda^2)/\pi\approx0.0526925$ and $c_2=\mathcal{E}(0,1,0)=(\ln2)/(\pi\lvert\Delta\rvert)$. The crossover from linear to constant behavior occurs at $\widetilde{T}_i=c_1/c_2$, i.e., at an initial temperature of $T_i\approx 0.239 (v/\gamma)^{1/2}$. This temperature also marks the transition from the initially adiabatic to the nonadiabatic cooling regimes in Figs.~\ref{f:E(t)}(b) and \ref{f:E(t)}(c).

Based on the homogeneity \eqref{e:homogeneous}, similar scaling laws can be obtained for $\mathcal{E}$ as a function of $\gamma/v$,
\begin{equation}\label{e:tildeE2}
T_i^{-1/z}\mathcal{E}(0,T_i,\gamma/v) = \mathcal{E}(0,1,\widetilde{T}_i^{s+1}),
\end{equation}
and also for ramps ending at nonzero temperatures; see Sec.~I of \cite{Note2}.

{\em Model-independent scaling theory.---}The derivation of the scaling relations \eqref{e:homogeneous} and \eqref{e:tildeE1}, while formally presented for the Kitaev chain in a thermalizing bosonic bath, is insensitive to many of the model's details. From a physical point of view, this is expected: It is characteristic for Kibble-Zurek physics that key results depend only on a few basic (and often universal) ingredients like the critical exponents of the underlying quantum phase transition, whereas specific details of the model do not play much of a role. 

From a technical point of view, the model-independence of our results can be understood from the derivation of the scaling relations in Sec.~II of the Supplemental Material \cite{Note2}: While a rate equation of the form \eqref{e:rateequation} is required, the precise functional form of the relaxation rate \eqref{e:rate} is not crucial. In fact, any relaxation rate that is an arbitrary function of $T/\lambda_k$, multiplied by some power of $\lambda_k$, will result in $\mathcal{E}(T,T_i,\gamma/v)$ being a generalized homogeneous function. Rate equations for mode occupation numbers appear widely for Markovian open quantum systems. They result generically from the class of quadratic thermalizing master equations considered in Ref.\ \cite{DAbbruzzoRossini21}, of which our setup is one example. Even for cases where a bath-induced coupling between the various excitation modes exists, as in Ref.\ \cite{Patane_etal08,*Patane_etal09}, the asymptotic approach to equilibrium is still believed to be well described by the rate equation picture, albeit with a single collective relaxation rate capturing the coupling between the excitation modes. The scaling behavior of this excitation rate will then impact the scaling of the excitation density itself, as confirmed in Ref.\ \cite{Patane_etal08,*Patane_etal09} for transverse-field Ising and $XY$ chains by direct calculations with Keldysh techniques.


{\em Discussion.---}A practical merit of temperature ramps in the context of Kibble-Zurek-type nonequilibrium physics is the emergence of clean scaling laws, as is evident from Eqs.\ \eqref{e:homogeneous}--\eqref{e:tildeE2} and Fig.~\ref{f:E(t)}(c). This is in contrast to parameter ramps at nonzero $T$, as illustrated by arrow (b) in Fig.~\ref{f:rampingdiagram}, where the excitation density $\mathcal{E}$ is a sum of two different power laws in the limit of weak system--bath coupling \cite{Patane_etal08,Patane_etal09}, and even more complicated otherwise. More complicated functional forms restrict the observation of universal scaling laws to narrow parameter regimes and render it challenging, if not impossible, to extract critical exponents from numerical or experimental data. Similarly, no clean scaling laws are obtained for cooling protocols at fixed noncritical values $\mu\neq\mu_c$; see Sec.~IV of \cite{Note2}.

While temperature ramps give rise to clean scaling laws containing the dynamical critical exponent $z$, no other critical exponent of the transition features. Parameter ramps at $T>0$, on the other hand, are also influenced by the correlation length critical exponent $\nu$, but not in the form of clean scaling laws. A solution to this issue, i.e., a strategy for ``cleaning up'' polluted scaling laws while retaining a $\nu$ dependence, may again be based on $T$-dependent protocols: Simultaneous ramping of $T$ and a suitable power of the Hamiltonian parameter $\mu$ is expected to produce clean scaling laws containing a combination of the exponents $z$ and $\nu$, hence providing information complementary to that obtained through pure cooling protocols.

{\em Conclusions.---}Cooling a system toward its quantum critical point gives rise to universal nonequilibrium scaling behavior governed by equilibrium quantum critical exponents. We established the occurrence of scaling laws based on an exact solution of the Lindblad equation describing a Kitaev chain coupled to bosonic baths, and subsequently argued that similar scaling laws hold quite generally whenever mode occupations are governed by rate equations of the form \eqref{e:rateequation}.

We presented examples where the temperature is ramped all the way to $T=0$, but this is not mandatory: The homogeneity \eqref{e:homogeneous} of the excitation density with respect to $T$ implies that ramps ending at suitable, small but positive, temperatures likewise lead to scaling behavior, which is a more realistic scenario for applications. Moreover, it may be useful for applications to replace linear temperature ramps by power laws $T(t)=T_i(1-vt/T_i)^\eta$, which lead to scaling with modified exponents, $\mathcal{E}\sim(v/\gamma)^{1/[z(s+1/\eta)]}$. Experimental verifications of our findings may be possible, in principle, in any of the numerous solid-state realizations of quantum phase transitions, conditional on the feasibility of precise temperature control and experimental accessibility of mode occupation numbers. Alternatively, recent proposals of cold atom-based analog quantum simulators at finite temperature promise increased control and flexibility \cite{Portugal_etal22,*MildenbergerMSc}, but still await experimental realization.

Our results open up avenues for generalizations and extensions. The effect of long-range interactions can straightforwardly be explored in open Kitaev chains, which remain analytically solvable in the presence of long-range hopping and/or pairing terms. Results are shown in Sec.~V of \cite{Note2} and reveal changes of the quantum critical exponents for sufficiently long-ranged hopping. Potential benefits of simultaneous ramping of $T$ and a Hamiltonian parameter $\mu$ have been outlined in a previous paragraph. Another interesting direction for future work, and also a further step toward applications, is the study of temperature ramps for models that couple to thermal baths only locally, for example at the ends of a chain, which results in temperature gradients that will alter the creation of excitations. 

\acknowledgments
M.\,K.\ acknowledges helpful discussions with Giovanna Morigi and Guido Pupillo during the early stages of this project, and with Nicolò Defenu on the long-range Kitaev chain. E.C.K.\ acknowledges financial support by the National Institute for Theoretical Physics of South Africa through a Master of Science bursary.

{\em Note added:} Related results for the transverse-field Ising model have recently been reported in Ref.~\cite{BacsiDora}.

\bibliography{../../../../MK.bib}


\end{document}


\title{Supplemental Material for\texorpdfstring{\\}{} ``Universal Cooling Dynamics Toward a Quantum Critical Point''}

\author{Emma C. King}
\affiliation{Institute of Theoretical Physics, University of Stellenbosch, Stellenbosch 7600, South Africa}

\author{Johannes N. Kriel}
\affiliation{Institute of Theoretical Physics, University of Stellenbosch, Stellenbosch 7600, South Africa}

\author{Michael Kastner}
\affiliation{Institute of Theoretical Physics, University of Stellenbosch, Stellenbosch 7600, South Africa}
\affiliation{Hanse-Wissenschaftskolleg, Lehmkuhlenbusch 4, 27753 Delmenhorst, Deutschland}

\date{\today}

\maketitle 

\onecolumngrid

\section{Scaling of the excitation density \texorpdfstring{$\mathcal{E}$}{E} as a function of the ramp rate \texorpdfstring{$v$}{v}}

\begin{figure*}\centering
\hfill
\includegraphics[width=0.49\linewidth]{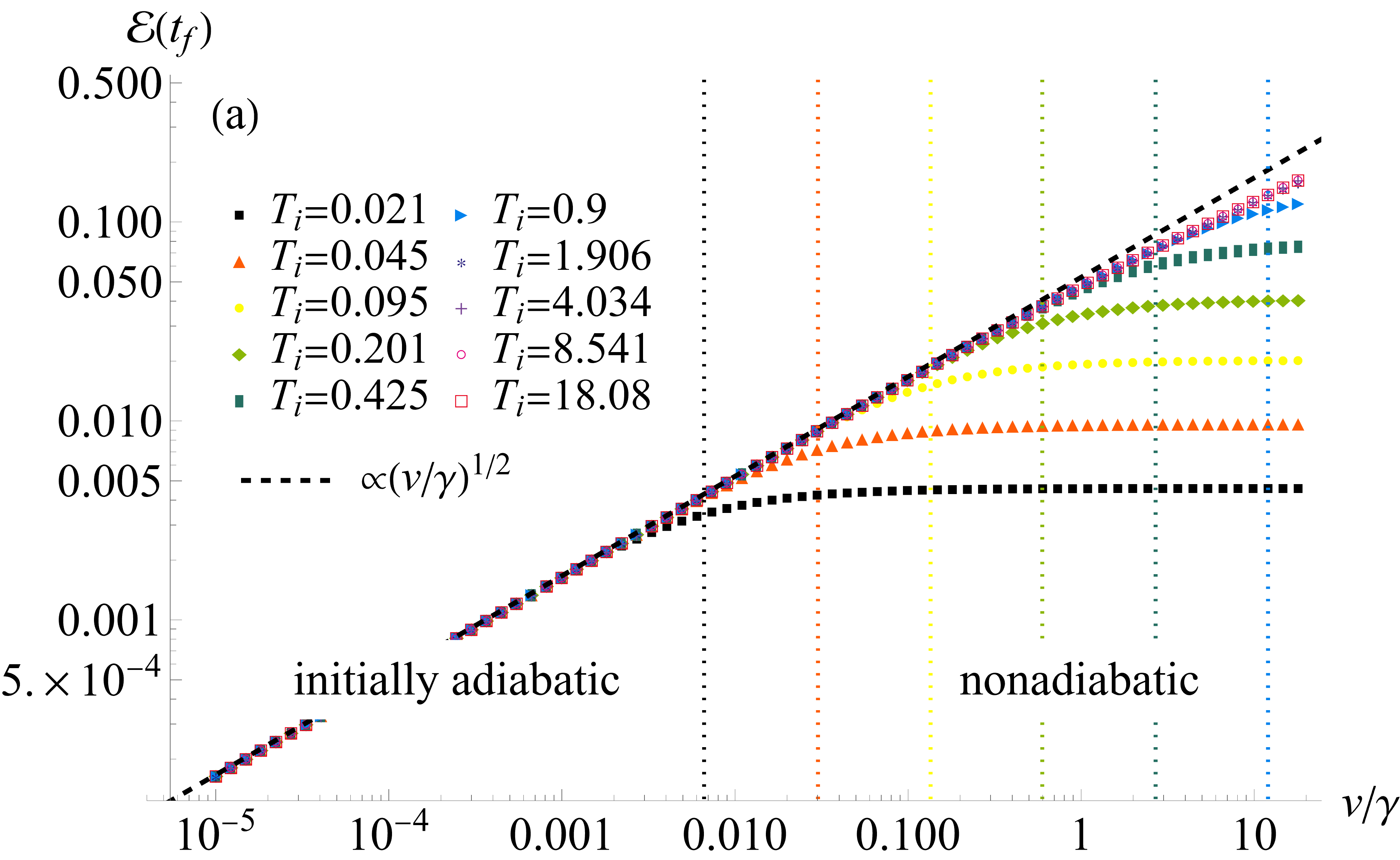}
\hfill
\includegraphics[width=0.49\linewidth]{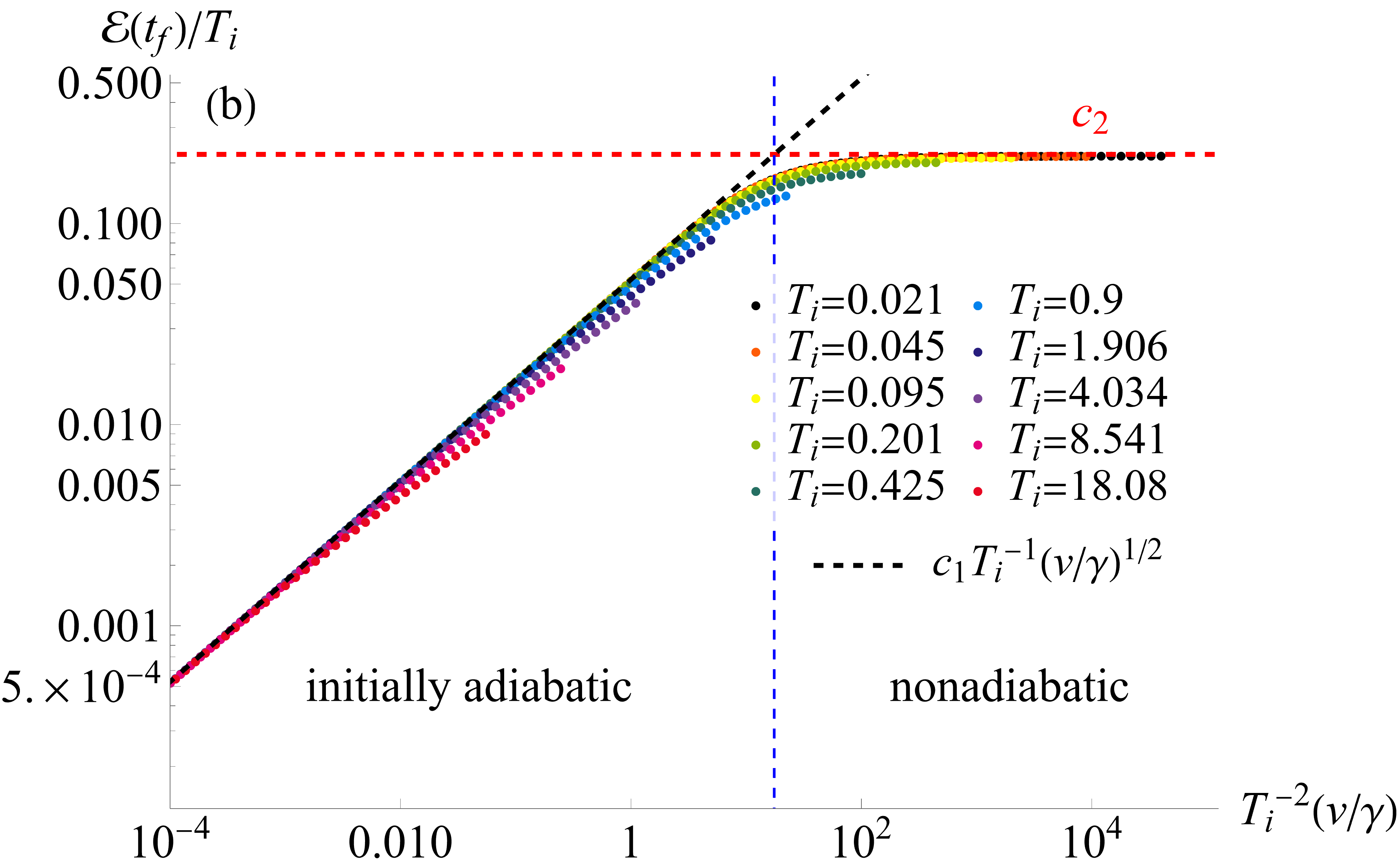}
\hfill\vphantom{.}
\caption{\label{f:E_vs_v} (a) The excitation density $\mathcal{E}$ at the end of the ramping protocol as a function of $v/\gamma$, shown for various initial temperatures $T_i$. Parameter values are $L=4\times 10^4$, $J=\Delta=1$, $\mu=-1$. Dotted vertical lines mark the crossover values between adiabatic and nonadiabatic regimes. The dashed black line shows the line $\mathcal{E}\propto \sqrt{v/\gamma}$ on which the data come to lie asymptotically for $T_i\to\infty$, in which case the dynamics is initially adiabatic for arbitrarily large ramp rates. (b) As in (a), but showing the rescaled excitation density $\mathcal{E}T_i^{-1}$ {\em vs.}\ the rescaled ramp rate $(v/\gamma)T_i^{-2}$, which results in approximate data collapse onto a single curve. For larger values of $T_i$, the quality of the data collapse diminishes (not shown).
}%
\end{figure*}

In Figs.~2(b) and 2(c) of the main text we show the excitation density $\mathcal{E}$ at the end of the linear cooling protocol, plotted as a function of the initial temperature $T_i$ of the ramp. As our subsequent scaling analysis reveals, $T_i$ is not the only scaling variable. In Fig.~\ref{f:E_vs_v} we show plots of $\mathcal{E}$ at the end of the protocol {\em vs.}\ the ramp rate $v$, confirming the scaling behavior with respect to that latter variable for the Kitaev chain: Plotting of the rescaled excitation density $\mathcal{E}T_i^{-1}$ {\em vs.}\ the rescaled ramp rate $(v/\gamma)T_i^{-2}$ results in data collapse onto a single curve to good approximation.

\section{Derivation of Eqs.~(6)--(7\texorpdfstring{\lowercase{b}}{b}) of the main text}
The starting point of this derivation is the rate equation (3) of the main text with a relaxation rate as in Eq.\ (4). We assume that, as it should, the cutoff frequency of the bath spectral density is chosen sufficiently large such that $\mathcal{J}(\lambda_k)=\pi\delta\lambda_k^s$ holds for the physical situation under investigation. Changing to the new variable $x=T(t)/\lambda_k=(T_i-vt)/\lambda_k$ in the rate equation (4) we obtain
\begin{equation}\label{e:firstOrderLinearDERateEqTempRamps1}
\frac{d\mathcal{P}_k}{dx} = \frac{2\pi\delta\gamma\lambda_k^{s+1}}{v}\coth[1/(2x)] \left[\mathcal{P}_k-\mathcal{P}_k^\text{th}(1/x)\right],
\end{equation}
which is a linear, first-order differential equation with nonconstant coefficients. According to Sec.~1.1.4 of Ref.\ \cite{PolyaninZaitsev}, the differential equation
\begin{equation}
\frac{dy(x)}{dx} = f_1(x)y(x) + f_0(x)
\label{e:REgeneralSol1}
\end{equation}
with arbitrary functions $f_0$ and $f_1$ has the solution
\begin{equation}
 y(x) = C e^{F(x)} + e^{F(x)}\int dx\, e^{-F(x)}f_0(x) \qquad\text{with}\qquad F(x) = \int{dx f_1(x)},
\label{e:REgeneralSol2}
\end{equation}
where the constant $C$ is determined by the initial condition. Identifying $y$ with $\mathcal{P}_k$, $f_1(x)$ with $2\pi\delta\gamma\lambda_k^{s+1}v^{-1}\coth[1/(2x)]$, and $f_0(x)$ with $-f_1(x)\mathcal{P}_k^\text{th}(1/x)$, and furthermore fixing the integration constant and domain such that $\mathcal{P}_k(T_i/\lambda_k)=\mathcal{P}^\text{th}(\lambda_k/T_i)$, one obtains the excitation probability of the $k$th mode,
\begin{equation}
\mathcal{P}_k(x) = Ce^{F(x)} - \frac{2\pi\delta\gamma\lambda_k^{s+1}}{v} e^{F(x)}\int_{T_i/\lambda_k}^{x} dx' e^{-F(x')}\coth[1/(2x')]\mathcal{P}^\text{th}(1/x'),
\label{e:excitationProbInTermsOfIntegralsTempRamps1}
\end{equation}
with
\begin{equation}
C = \mathcal{P}^\text{th}(\lambda_k/T_i) \qquad\text{and}\qquad F(x) = \frac{2\pi\delta\gamma\lambda_k^{s +1}}{v} \int_{T_i/\lambda_k}^{x} dx''\coth[1/(2x'')].
\label{e:excitationProbInTermsOfIntegralsTempRamps2}
\end{equation}
For later analyses it is useful to make explicit the dependence of $\mathcal{P}_k(T/\lambda_k)$ on the various parameters and on the initial temperature. To this end, we identify $\mathcal{P}_k(T/\lambda_k)$ with the trivariate function
\begin{equation}
\mathcal{P}\left(\frac{T}{\lambda_k},\frac{T_i}{\lambda_k},\frac{\gamma\lambda_k^{s +1}}{v}\right)\equiv \mathcal{P}_k(T/\lambda_k),
\end{equation}
for which the expression following from Eqs.~\eqref{e:excitationProbInTermsOfIntegralsTempRamps1} and \eqref{e:excitationProbInTermsOfIntegralsTempRamps2} appears in (7a) and (7b) of the main text. 
%
%
The excitation density (5) is defined as a sum over the mode occupations $\mathcal{P}_k$, which, in the limit of large system size, can be replaced by the integral
\begin{equation}\label{e:densityOfExcitationDefinitionChap5p2Nr2}
\mathcal{E}(T,T_i,\gamma/v) = \frac{1}{\pi}\int_{0}^{\pi}{dk\, \mathcal{P}\left(\frac{T}{\lambda_k},\frac{T_i}{\lambda_k},\frac{\gamma\lambda_k^{s +1}}{v}\right)},
\end{equation}
where we have exploited the $k\leftrightarrow-k$ mode symmetry to halve the integration domain. At the quantum critical point the energies of the low-lying modes are given, to leading order in $k$, by $\lambda_k = c|k|^z$ \cite{Dutta_etal} with $c$ a constant and $z$ the equilibrium dynamical critical exponent. Within this low-energy approximation, we extend the upper bound of the integral in Eq.~\eqref{e:densityOfExcitationDefinitionChap5p2Nr2} to infinity and perform the change of variables $k\rightarrow (\lambda/c)^{1/z}$, which yields Eq.\ (6) of the main text.

\section{Asymptotic behavior of the excitation density in Eqs.~(6)--(7\texorpdfstring{\lowercase{b}}{b}) of the main text}
\label{s:asymptotics}

To establish the asymptotic behavior of $\mathcal{E}(0,T_i,\gamma/v)$ in the limit of large $T_i$, we prove below that
\begin{equation}\label{e:c1}
\lim_{y\to\infty}\mathcal{E}(0,y,1)=c_1,
\end{equation}
with $c_1$ a finite, nonzero constant. Then it follows from Eq.\ (9) of the main text that
\begin{equation}
\lim_{T_i\to\infty}\mathcal{E}(0,T_i,\gamma/v)=(\gamma/v)^{-1/[z(s+1)]} c_1,
\end{equation}
which accounts for the plateaux in Figs.~2(b) and 2(c) of the main text. 
To establish the limit in Eq.~\eqref{e:c1} we use Lebesgue's dominated convergence theorem. First we apply integration by parts in Eq.~(7a) to write 
\begin{equation}
	\mathcal{P}\left(0,y/\lambda,\lambda^{s+1}\right)=\int_0^{y/\lambda} dx'\, \left[\frac{d}{dx'} \mathcal{P}^\text{th}(1/x')\right] \exp\left[-\lambda^{s+1}f(x',0)\right].
	\label{e:ExcitationDensityAfterIBP}
\end{equation}
From Eq.\ (7b) it follows that $f(x',0)$ is positive and quadratically increasing for large $x'$, hence the factor $\exp[-\lambda^{s+1}f(x',0)]$ is exponentially damped in $x'$. The term in square brackets in Eq.~\eqref{e:ExcitationDensityAfterIBP}, which is a derivative of the Fermi-Dirac distribution, is positive and finite on the domain of integration and tends to zero for large $x'$. As a result, the integral of the product of the two terms converges and is finite also in the limit of the domain of integration extending to $+\infty$. Since the integrand is positive it holds that $0<\mathcal{P}\left(0,y/\lambda,\lambda^{s +1}\right)<\mathcal{P}(0,\infty,\lambda^{s+1})$. Next we seek a bound on $\mathcal{P}(0,\infty,\lambda^{s+1})$. Returning to the integrand in Eq.~\eqref{e:ExcitationDensityAfterIBP}, we note that since $f(x',0)\geqslant 2\pi\delta x'$ it holds that $\exp[-\lambda^{s+1}f(x',0)]\leqslant \exp(-2\pi\delta \lambda^{s+1} x')$, and also that the derivative in the square brackets is bounded from above by $\exp(-1/x')/x'^2$. Together, this allows us to bound the integrand of $\mathcal{E}(0,y,1)$ in Eq.~\eqref{e:c1}, see also Eq.~(6), as
\begin{equation}
	0\leqslant \lambda^{1/z-1} \mathcal{P}\left(0,y/\lambda,\lambda^{s +1}\right) \leqslant \lambda^{1/z-1} \mathcal{P}\left(0,\infty,\lambda^{s+1}\right) \leqslant \lambda^{1/z-1} \int_0^\infty dx'\, \frac{1}{x'^2}e^{-1/x'} \exp(-2\pi\delta \lambda^{s+1} x') \equiv U(\lambda).
\end{equation}
Using Eq.~10.32.10 from Ref.~\cite{NIST} we find that $U(\lambda)=2\lambda^{1/z-1}\sqrt{b}K_1(2\sqrt{b})$, with $b=2\pi\delta \lambda^{s+1}$ and $K_1$ a modified Bessel function of the second kind. Note that $z\leqslant 1$ in all the cases we consider. The asymptotic behavior of $K_1$ in Eq.~10.40.2 of Ref.~\cite{NIST} now implies that $U$ behaves as $U(\lambda)\sim \lambda^{1/z-1}\sqrt{\pi}b^{1/4}e^{-2\sqrt{b}}$ in the limit of large $\lambda$. It follows that the integral $\int_0^\infty d\lambda\, U(\lambda)$ converges, which then implies the existence of the limit in Eq.~\eqref{e:c1}.

The asymptotic behavior of $\mathcal{E}(0,T_i,\gamma/v)$ in the limit of small $T_i$ can be established by proving that
\begin{equation}
\lim_{y\searrow0}\mathcal{E}(0,1,y)=c_2,
\end{equation}
with $c_2=\ln2/(\pi\lvert\Delta\rvert)$, which follows when inserting $\mathcal{P}(0,1/\lambda,0)=[\exp(\lambda)+1]^{-1}$ into Eq.\ (6) of the main text and setting $z=1$. Then it follows from Eq.\ (10) of the main text that
\begin{equation}
\lim_{T_i\searrow0}\mathcal{E}(0,T_i,\gamma/v)=T_i^{1/z} c_2,
\end{equation}
which, when inserting the dynamical critical exponent $z=1$ of the Kitaev chain, accounts for the linear increases on the left-hand sides of Figs.~2(b) and 2(c) of the main text.

By similar techniques one can analyze the asymptotic behavior of $\mathcal{E}$ as a function of $v$, and the constant, respectively linear, regimes in Fig.\ \ref{f:E_vs_v} are recovered and characterized.

\begin{figure*}\centering
\hfill
\includegraphics[width=0.49\linewidth]{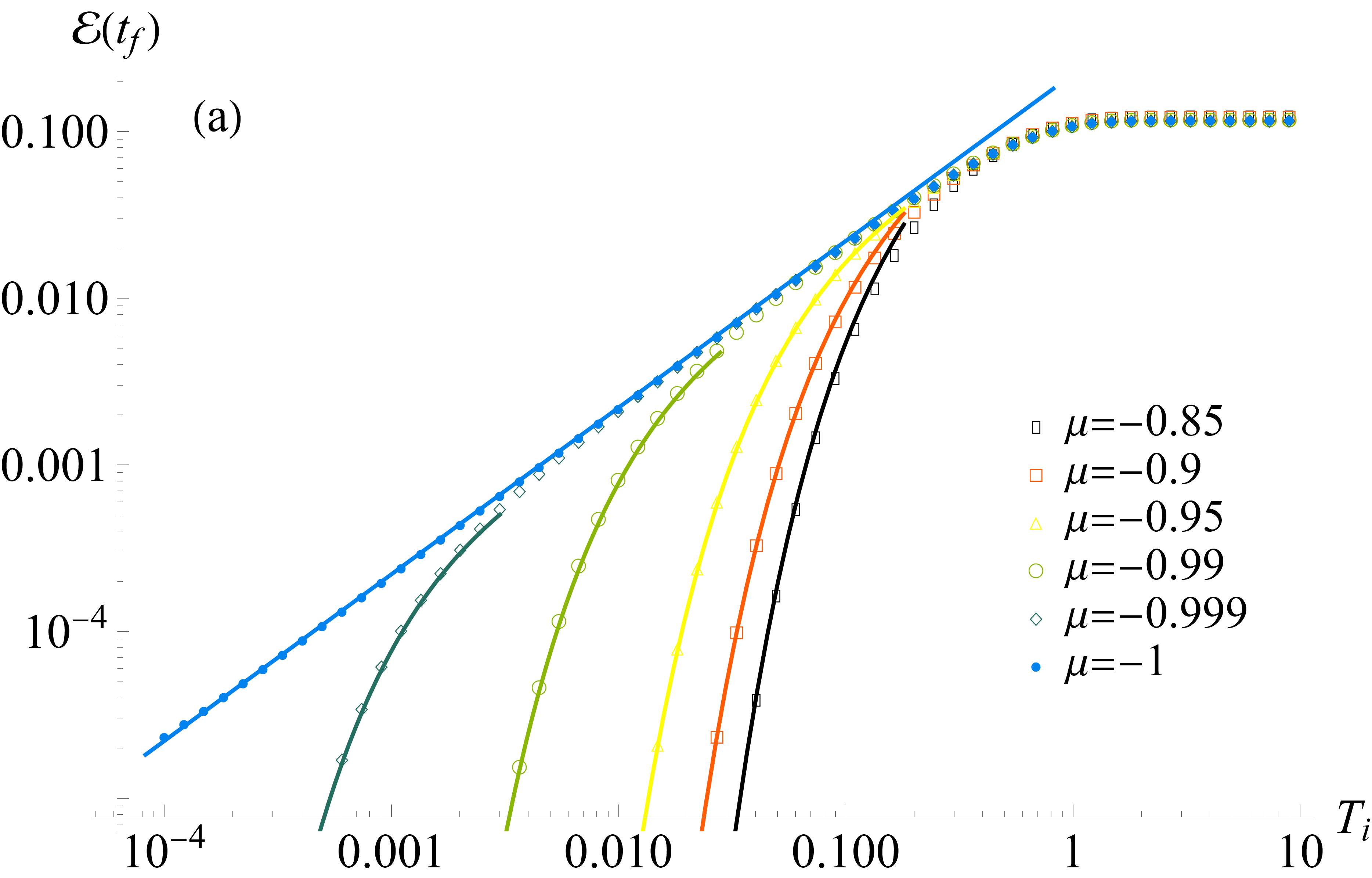}
\hfill
\includegraphics[width=0.49\linewidth]{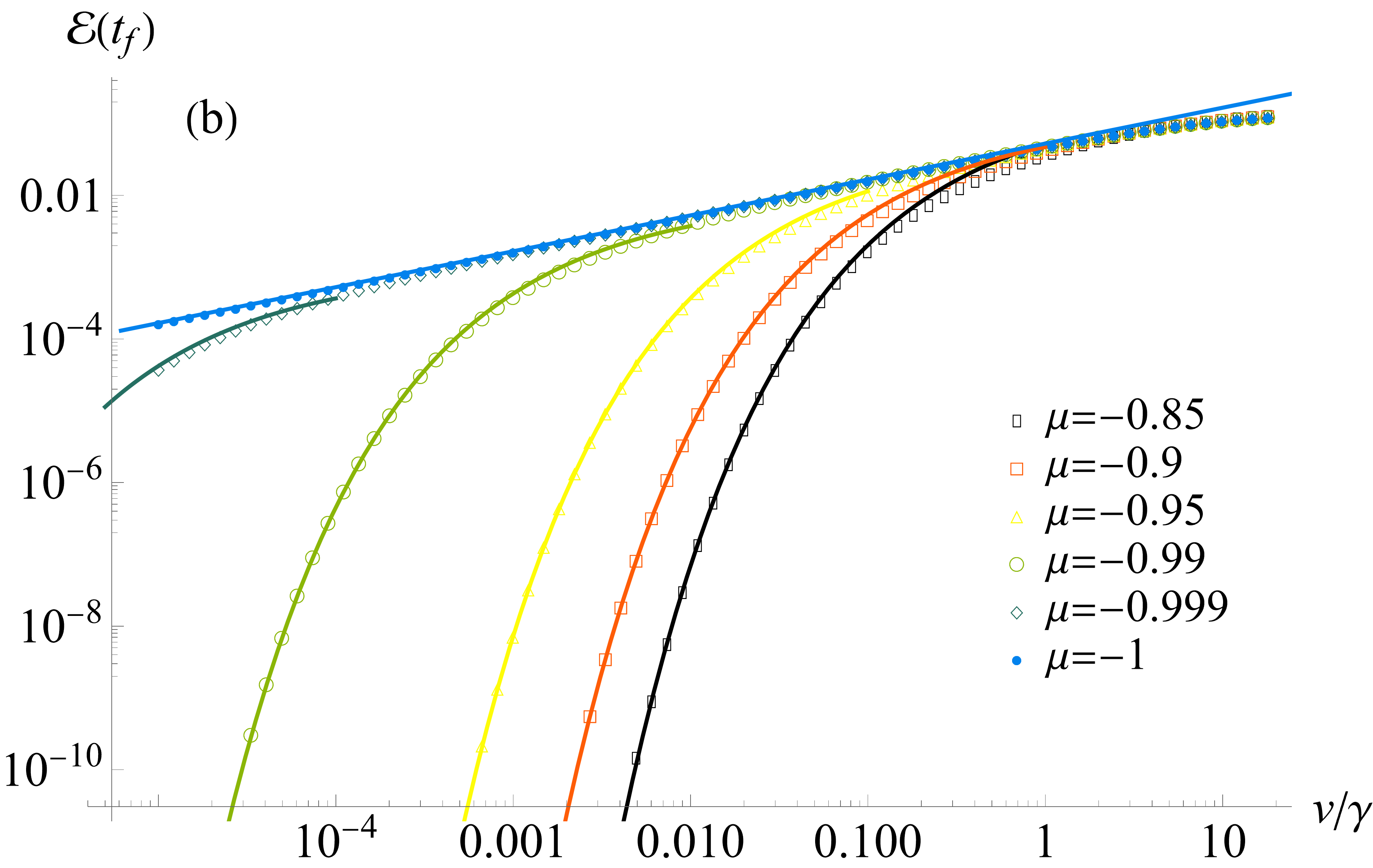}
\hfill\vphantom{.}
\caption{\label{f:ED_away_from_CP} (a) The final excitation density $\mathcal{E}(t_f)$ as a function of the initial temperature $T_i$, shown for protocols at various fixed noncritical values of the chemical potential $\mu$. Parameter values are $L=4\times 10^4$, $J=\Delta=1$, $\delta=1$, and $v/\gamma=8.1$. The solid blue line corresponds to the scaling result $\mathcal{E}\propto T_i$ for the ramp at $\mu=\mu_c=-1$, as in Fig.~2 of the main text. The asymptotic result in Eq.~\eqref{e:scaling_away_from_CP_1} is shown for each value of $\mu\neq -1$ as a solid line in the corresponding color. (b) As in (a), but now showing $\mathcal{E}(t_f)$ as a function of $v/\gamma$ for $T_i=0.9$. The solid blue line shows the scaling behavior $\mathcal{E}\propto \sqrt{v/\gamma}$ for cooling at $\mu=\mu_c=-1$, as in Fig.~\ref{f:E_vs_v}. The asymptotic results of Eq.~\eqref{e:scaling_away_from_CP_2} for the cooling at $\mu\neq\mu_c$ are shown as solid lines in the corresponding colors.
}%
\end{figure*}

\section{Cooling protocols at fixed noncritical values of \texorpdfstring{$\mu$}{mu}}
In the main text we considered cooling protocols at $\mu=\mu_c$ in order to probe aspects of the quantum critical behavior in the cleanest possible way. For cooling protocols at fixed noncritical values $\mu\neq\mu_c$, the asymptotic low-$T_i$ behavior of the excitation density is found to be
\begin{equation} \label{e:scaling_away_from_CP_1}
	\mathcal{E}=\frac{1}{2}\sqrt{\frac{T_i}{\pi\lambda_1}}\exp\left(-\frac{\lambda_0}{T_i}\right),
\end{equation}
where $\lambda_k=\lambda_0+\lambda_1 k^2$ is the low-energy approximation of the mode energies away from the critical point. In particular, we have $\lambda_0=2|\mu+J|$ and $\lambda_1=(\Delta^2-4J(\mu+J))/(4|\mu+J|)$ for the short-range Kitaev chain, and we consider parameter choices for which $\lambda_1>0$. Similarly, the small $v/\gamma$ behavior of the excitation density is given by
\begin{equation} \label{e:scaling_away_from_CP_2}
	\mathcal{E}=\frac{1}{\sqrt{8}}\sqrt{\frac{\lambda_0}{\lambda_1}}\exp\left(-\sqrt{8\pi}\lambda_0\sqrt{\frac{\gamma}{v}}\right).
\end{equation} 
The derivations of these asymptotic expressions, which are based on the rate equation solution in Eqs.~(6)--(7b) of the main text, will be given elsewhere. In support of these results, Fig.~\ref{f:ED_away_from_CP} shows a comparison of the two asymptotic expressions for $\mathcal{E}$ with numeric data, where we plot the equivalents of Fig.~2(b) of the main text and Fig.~\ref{f:E_vs_v}(a), now for cooling protocols at fixed $\mu\neq\mu_c$. 
As Fig.~\ref{f:ED_away_from_CP} corroborates, the exponential factors in expressions \eqref{e:scaling_away_from_CP_1} and \eqref{e:scaling_away_from_CP_2} rule out any simple power-law scaling of $\mathcal{E}$ with either $T_i$ or $v/\gamma$ away from the critical point.


 \begin{figure}
\centering
 \includegraphics[width=0.55\linewidth]{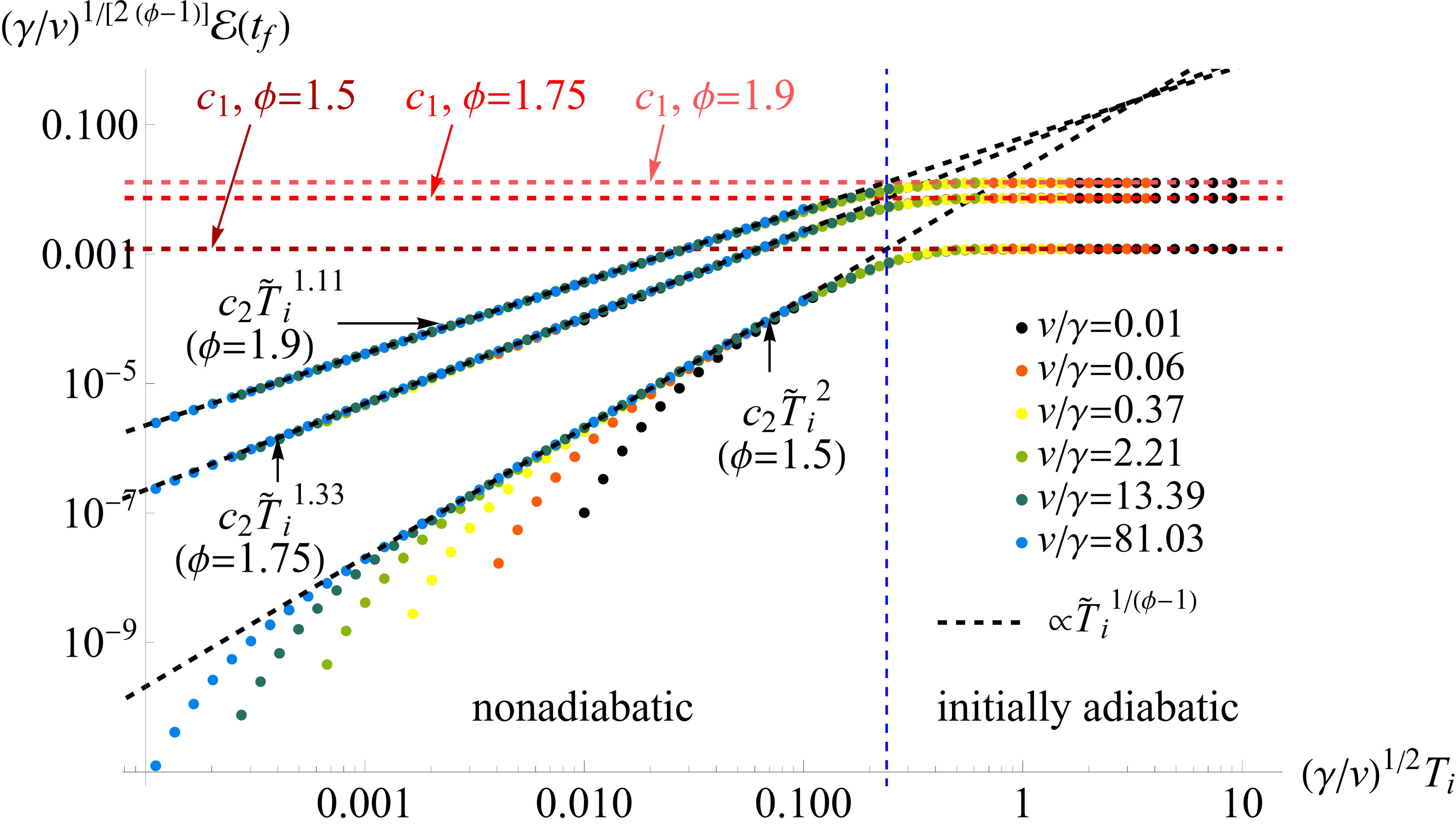}%
 \caption{Rescaled excitation density $(\gamma/v)^{1/(2z)}\mathcal{E}$ at the end of a temperature ramp (3) as a function of the rescaled initial temperature $\widetilde{T}_i=\sqrt{\gamma/v}T_i$ for the Kitaev chain with long-range hopping \eqref{e:LongRangeKitaev}. Data is shown for the long-range hopping exponents $\phi=1.5$, $\phi=1.75$ and $\phi=1.9$, and for various ramp rates in each case, as indicated in the legend. For each data set, the corresponding asymptotic behaviors in the small-$\widetilde{T}_i$, respectively large-$\widetilde{T}_i$, regimes are shown as red and black dashed lines. Parameter values used are $L=4\times 10^6$, $J=\Delta=1$, $\delta=1$, and $\mu=\mu_c=-\zeta(\phi)$ \cite{Vodola_etal16,KingKastnerKriel} given in terms of the Riemann zeta function $\zeta$.}%
 \label{fig:dataCollapseLRHopWithTi}%
 \end{figure}

\section{Nonequilibrium scaling of the long-range Kitaev chain}

The interaction range of an interacting many-body system, or the hopping and/or pairing range of a quasifree many-body quantum system, can strongly affect the system's properties. In particular, the critical behavior of long-range models, including the numerical values of critical exponents, usually differs from their short-range counterparts \cite{FisherMaNickel72,DuttaBhattacharjee01,Defenu_etal20}. Hence, we also expect the scaling laws (8)--(10) for the cooling ramps to be influenced by the range.

We consider a Kitaev chain of length $L$ with long-range hopping and pairing \cite{Vodola_etal16},
\begin{equation}\label{e:LongRangeKitaev}
			H = J \sum_{j=1}^L{\sum_{l=1}^{\left\lfloor L/2 \right\rfloor}{d_l^{-\phi}\left( c_j^\dagger c_{j+l}^{\vphantom{\dagger}} + c_{j+l}^\dagger c_j^{\vphantom{\dagger}}\right)}} + \frac{\Delta}{2}\sum_{j=1}^L{\sum_{l=1}^{\left\lfloor L/2 \right\rfloor}{d_l^{-\alpha}\left( c_j^{\vphantom{\dagger}} c_{j+l}^{\vphantom{\dagger}} + c_{j+l}^\dagger c_j^\dagger\right)}} + 2 \mu \sum_{j=1}^L{c_j^\dagger c_j^{\vphantom{\dagger}}},		
\end{equation}
with periodic boundary conditions. Different from the nearest-neighbor Hamiltonian (1) of the main text, hopping and pairing takes place between all pairs of sites. The corresponding hopping and pairing strengths decay like power laws proportional to $d_l^{-\phi}$ and $d_l^{-\alpha}$, respectively, where
\begin{equation}
d_l = \begin{cases}
l &\text{for $l<L/2$},\\
2^{1/\phi}l &\text{for $l = L/2$,}
\end{cases}
\end{equation}
is a distance along the chain that takes into account the periodic boundary conditions. The factor of $2^{1/\phi}$ above will only survive in the hopping terms, where it prevents an overcounting of terms with a hopping range of $l=L/2$. In the limit of the long-range exponents $\phi$ and $\alpha$ going to infinity, the Kitaev chain with nearest-neighbor hopping and pairing (1) is recovered.

Just like the nearest-neighbor version, the Kitaev chain with long-range hopping and pairing \eqref{e:LongRangeKitaev} is diagonalizable by means of a Bogoliubov transformation; see Ref.\ \cite{Vodola_etal16} and the companion paper \cite{KingKastnerKriel} for details. The presence of long-range terms leads to a richer phase diagram \cite{Viyuela_etal16,BhattacharyaDutta18}, and also to modified quantum critical exponents once $\phi$ or $\alpha$ become sufficiently small \cite{Vodola_etal16}. It is this latter property which we are going to probe by means of temperature ramps toward quantum criticality. The solvability of the long-range Kitaev chain, as in the nearest-neighbor case, also extends to the corresponding Lindblad equation describing the system coupled to bosonic baths, which can be diagonalized in Liouville space by the method of Third Quantization \cite{Prosen08}; see Ref.\ \cite{KingKastnerKriel} for details. 

In Fig.~\ref{fig:dataCollapseLRHopWithTi} we plot data equivalent to that of Fig.~2 of the main text, but for the Kitaev chain with long-range hopping for various values of the long-range exponent $\phi$. For this model the dynamical critical exponent is
\begin{equation}\label{e:z}
z=\begin{cases}
\phi-1 & \text{for $1\leqslant\phi\leqslant2$},\\
1 & \text{for $2\leqslant\phi$}.
\end{cases}
\end{equation}
According to the scaling law (9) of the main text, data collapse is obtained by plotting the rescaled excitation density $(\gamma/v)^{1/(2z)}\mathcal{E}$ at the end of a temperature ramp as a function of the rescaled initial temperature $\widetilde{T}_i=\sqrt{\gamma/v}T_i$. Hence, according to \eqref{e:z}, for $\phi\in[1,2]$ the rescaling has to be performed with a $\phi$-dependent exponent, which is nicely confirmed by Fig.~\ref{fig:dataCollapseLRHopWithTi}. According to the scaling laws of the main text, also the slope of the asymptotic behavior on the left-hand side of the plot becomes $\phi$-dependent in that case, $(\gamma/v)^{1/(2z)}\mathcal{E}\propto\widetilde{T}_i^{1/(\phi-1)}$, which again is confirmed by the data of Fig.~\ref{fig:dataCollapseLRHopWithTi}. The deviation from the asymptotic behavior at low initial temperatures is due to finite-size effects. Similar results are obtained for the case of long-range pairing (not shown). Note that the nonequilibrium scaling of temperature ramps in the long-range Kitaev chain is governed by the equilibrium critical exponents of the underlying quantum phase transitions, even in the regime where deviations from such behavior were found for parameter ramps \cite{Defenu_etal19}.

\bibliography{../../../../MK.bib}